\documentclass[aps,prl]{revtex4-2}

\usepackage{graphicx}
\usepackage{amsmath,amssymb}
\usepackage{hyperref}

\begin{document}

\title{Comment on \\'Anomalies in the electronic stopping of slow antiprotons in LiF' , arXiv:2501.14381}

\date{April 2025}

\author{G. Schiwietz}%
\affiliation{Helmholtz-Zentrum Berlin für Materialien und Energie GmbH,  
 Albert-Einstein-Str. 15, 12489, Berlin, Germany 
}%

\author{P.L. Grande}

\affiliation{Instituto de F\'isica da Universidade Federal do Rio Grande do Sul, Av. Bento Gon\c{c}alves, Porto Alegre, 9500, Brazil
}%
\maketitle

\section{Abstract}
This work contains detailed discussions on the contents 
of Phys. Rev. Lett. 134~\cite{PRL134}, in the following denoted PRL134. 
In this comment, we revisit and elaborate on the Adiabatic Ionization Model (AIM) for the energy loss of antiparticles in matter, with particular reference to its theoretical foundation as established in Refs.\cite{FermiTeller, MCSCF, JPB-HMI}. The AIM framework plays a central role in describing the ionization dynamics in the low-velocity regime considered by the authors of Ref.\cite{PRL134}. Calculated AIM results for the energy loss of $\bar p$ in LiF crystals are compared to experimental data and different other models, pointing to severe problems of the PRL134 results.  

Beyond this specific comparative theoretical investigation, we critically examine several statements and assumptions made in Ref.~\cite{PRL134}. 
Certain claims presented therein appear to be inconsistent with established theoretical principles or 
are insufficiently justified by the data and arguments provided. As such, we believe that further clarification, 
and in some cases, a more rigorous justification, is necessary to substantiate those points.

\section{The Adiabatic Ionization Model applied to $\rm \bf \bar p$ in solid $\rm \bf LiF$}
\label{sec:energy}

In the following, we discuss the local adiabatic energy gap of LiF in the presence 
of an additional particle (e.g. a proton or an antiproton) inside the bulk. 
These results deviate from the well known electronic density of states 
of an undisturbed solid, because an additional charged particle 
breaks the crystal symmetry and introduces a local quasi-atomic character
of the electronic properties. Starting from this adiabatic picture, 
we introduce a simple consideration of the effect of the $\bar p$ velocity 
inside LiF and discuss different theoretical energy loss results in comparison 
to the experimental $\bar p$ data that have been measured at CERN~\cite{PRL93}.

\subsection{Adiabatic Energy Gaps in LiF}
\label{subsec:gaps}

\begin{figure}[htb!]
    \centering
    \includegraphics[width=0.8\linewidth]{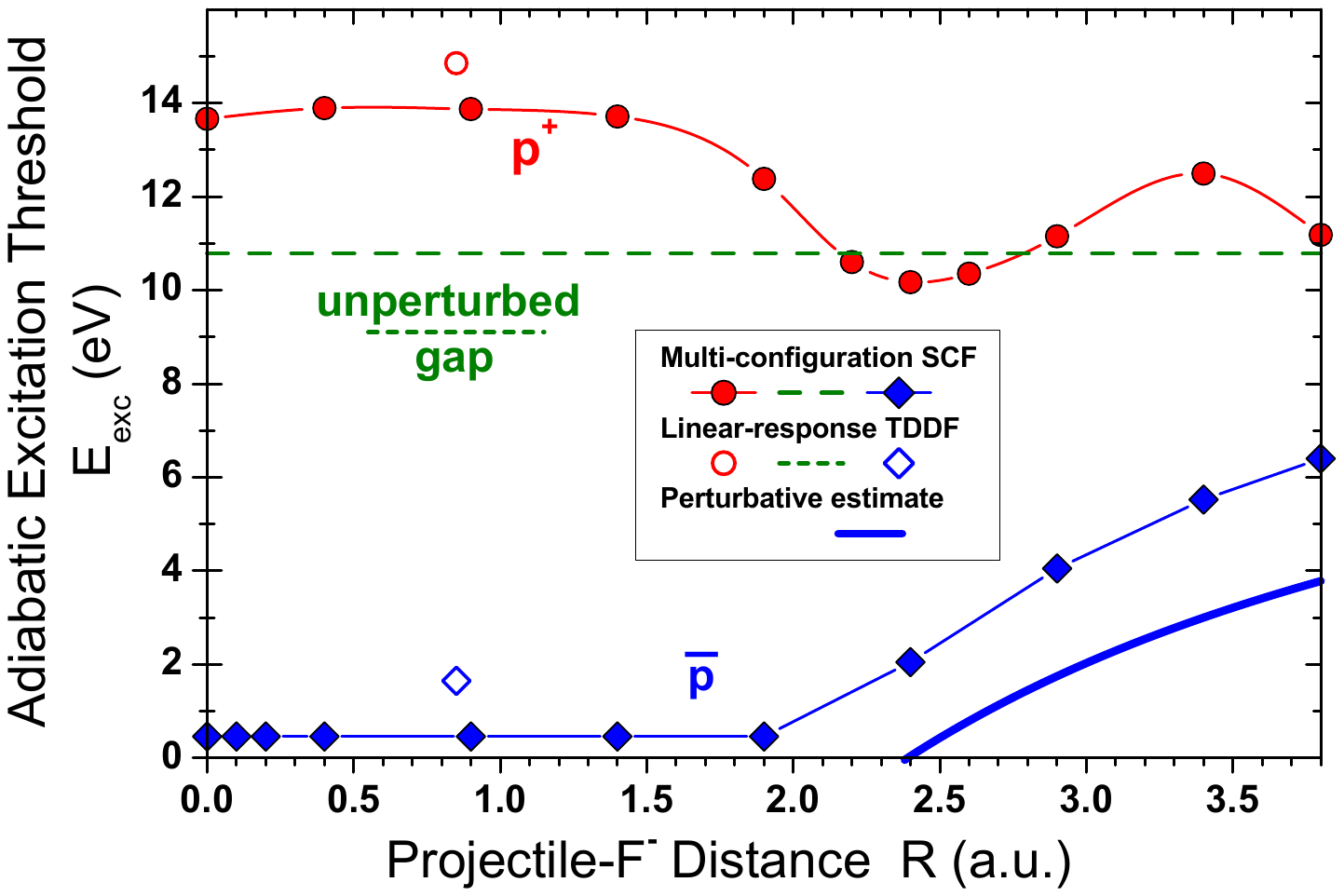}
    \caption{Value of the static excitation-energy gap vs. distance between a point charge 
$q$ (with $q= +e,~0,~or~-e$) and an $F^-$ ion of a LiF crystal (see text)}
    \label{fig:F1}
\end{figure}

Fig. \ref{fig:F1} displays the electronic excitation threshold of solid LiF resulting from different models for different physical situations. 
The dashed green lines show the unperturbed LiF gap. They have been obtained from a 
Multi-Configuration Self-Consistent Field solution (MCSCF, long-dashed line~\cite{MCSCF}) 
which is assumed to be very precise as well as from a linear-response version of 
a Time-Dependent Density-Functional approach (TDDF, short-dashed line~\cite{PRL128}). 
Recent experimental gap determinations range from $E_{gap}$=13.0 to 13.6~eV ~\cite{expGap}.
However, the main point of this plot is the discussion of what happens if a singly charged heavy particle is placed adiabatically inside the ionic crystal,  at a distance $R$ from the neighboring $F^-$ ion.  Occupied 2p orbitals near $F^-$ determine the excitation gap and are most sensitive to such a perturbation by either a proton (red curves and symbols) or an antiproton (blue curves and symbols). Although technically very different, both theories yield similar results. 

In the literature, there is agreement about the behavior of the energy loss of protons in LiF. In the light of PRL134 \cite{PRL134}, 
this seems not to be the case for $\bar p$ in LiF. Thus, from this point on, we focus exclusively on the antiproton results. It is seen in Fig. \ref{fig:F1} that the electronic excitation threshold of LiF is significantly reduced in the presence of an antiproton. The smaller the distance 
between $\bar p$ and $F^-$ ion, the lower is the minimum excitation energy. As becomes clear from SCF calculations~\cite{MCSCF}, this is mainly related to a strong positive energy shift (a promotion) of a partly occupied F-2p orbital. Such an energy shift was first predicted for insulators by Fermi and Teller~\cite{FermiTeller}. This shift for atomic hydrogen  was later confirmed within advanced quasi-molecular-orbital 
treatments and calculated also for other dipolar collision systems, such as $\bar p$ +H$_2$ and $\bar p$ +He ~\cite{VSM,JPB-HMI}. 
 
The solid blue curve in Fig. \ref{fig:F1} stands for an approximate first-order solution of such an antiproton-induced (dipolar) 
energy shift. Neglecting polarization and blow-up of the electron densities in the vicinity of the antiproton and keeping in mind that continuum energies in an infinite bulk material are not very sensitive to a local perturbation, we have simply considered the diagonal interaction matrix element of the uppermost $F^-(2p)$ orbital, using
\begin{equation}
    E_{exc}(eV) = E_{gap} - \langle 2p|1/(\vec R -\vec r)|2p \rangle
                \approx E_{gap} - 1/(R^2+r^2_{mean})^{0.5},
    \label{eq:diagonal}
\end{equation}
with $r_{mean} = 1/\langle2p|1/r|2p\rangle = 0.808 a.u.$, obtained from a self-consistent Hartree-Fock-Slater solution for atomic $F^-$.   Except for a distance shift of about 25\%, this curve agrees reasonably well with the advanced MCSCF results in the plot and with quasi-molecular-orbital results for an antiproton near atomic $F^-$, calculated with the code provided by U.Wille~\cite{VSM}. 
Thus, at large distances the repulsive interaction between $\bar p$ and bound electron leads 
to a coordinate-space polarization (the so-called Barkas effect), but at smaller distances, bound 
electrons can also be promoted significantly in energy. This promotion seems to be dominated 
by the reduced potential electron energy in the vicinity of the antiproton.
Considering MCSCF basis-set and cluster-size effects~\cite{MCSCF},  those results are consistent with a vanishing excitation gap leading to ionization of a single electron at distances $R <2~a.u.$.This is a direct confirmation of the estimate by Fermi and Teller.

\subsection{Collisional Broadening and Antiproton Energy Losses}
\label{subsec:Broadening}

\begin{figure}[htb!]
    \centering
    \includegraphics[width=0.8\linewidth]{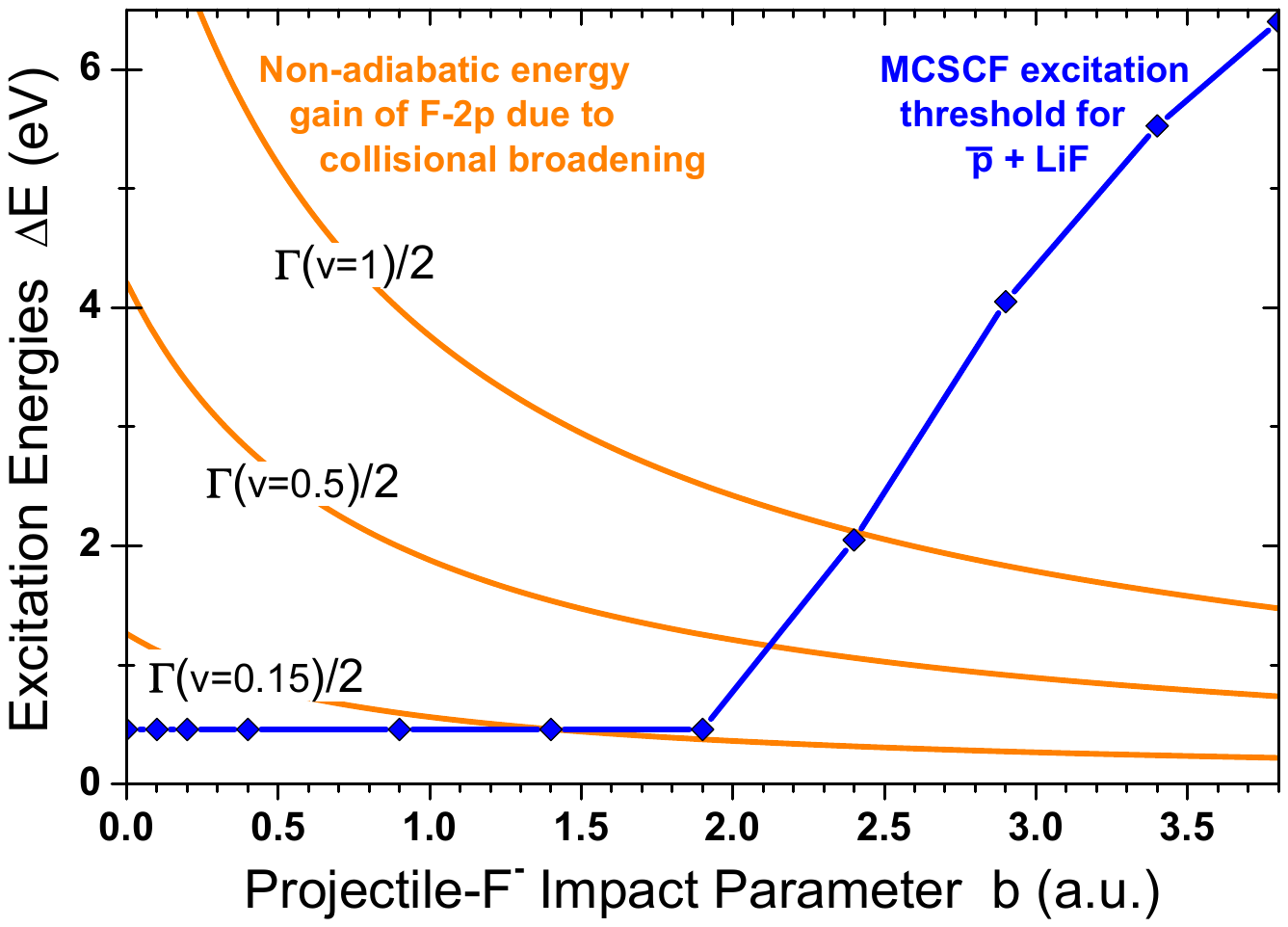}
    \caption{Scheme for extracting the impact-parameter threshold $b_c$ for electronic excitations at a fixed antiproton velocity $v$}
    \label{fig:F2}
\end{figure}

Fig. \ref{fig:F2} shows the MCSCF results~\cite{MCSCF} at the distance-of-closest-approach 
(or the corresponding impact parameter) together with an estimated 
collisional broadening for projectile speeds of 0.15, 0.5 and 1 a.u.. 
This broadening relates the finite interaction time $\tau$ (using Bohr's  prescription~\cite{JPB-HMI} for the characteristic interaction length) 
with an electronic energy spread $\Gamma$. A crossing of $\Gamma /2$ with the blue excitation curve indicates that the mean non-adiabatic energy fluctuations are equal to the adiabatic threshold and significant inelastic energy losses take place below the corresponding critical impact parameter $b_c$. 
This step-function assumption enables a simple estimation of energy loss cross sections in LiF. The specific energy loss is then derived via a simplified integral 
over impact parameters that may be approximated as~\cite{JPB-HMI}
\begin{equation}
    S^{\bar p}_{e}(v) = \pi ~P_{cont} ~\Delta E_{trans} ~ b_c^2.
    \label{eq:SeFormula}
\end{equation}
This defines the Adiabatic Ionization Model (AIM).\cite{JPB-HMI} For $\bar p$ in LiF, we have set the energy transfer $\Delta E_{trans}$ equal to the experimental (unperturbed) gap energy plus $1~eV$ as a mean excess-energy of ionized electrons ($\Delta E_{trans} = 14.5 eV$). Further, a mean ionization suppression factor of $P_{cont} \approx 58\%$ for the lowest continuum energies above threshold was estimated from calculated density-of-states~\cite{eDOS0,eDOS1}. It is emphasized that the concept of collisional broadening has been used successfully over decades of intense research on quasi-molecular excitation, ionization and energy-loss processes for ion-atom collisions at low projectile speeds (below 1 a.u.)~\cite{ArArMO}.  

Note that the step-function for the collisional broadening might be replaced by a more advanced treatment based on the  Weizs\"{a}cker-Williams method of virtual quanta~\cite{Jackson}  for generating an excitation frequency spectrum and a related excitation probability as a function of the energy gain.  Also, it would be possible to consider the impact-parameter dependence of the excess kinetic energy $\Gamma /2-E_{exc}$ (instead of just a mean value) to compute more accurate values of $\Delta E$ as well as $P_{cont}$. 
As an alternative ansatz for both effects, one might instead use the time-dependent perturbed-stationary-state theory~\cite{PSS-Basbas}.

\subsection{Energy Loss Comparison for Antiprotons at low velocities}
\label{subsec:Comparison}

\begin{figure}[htb!]
    \centering
    \includegraphics[width=0.65\linewidth]{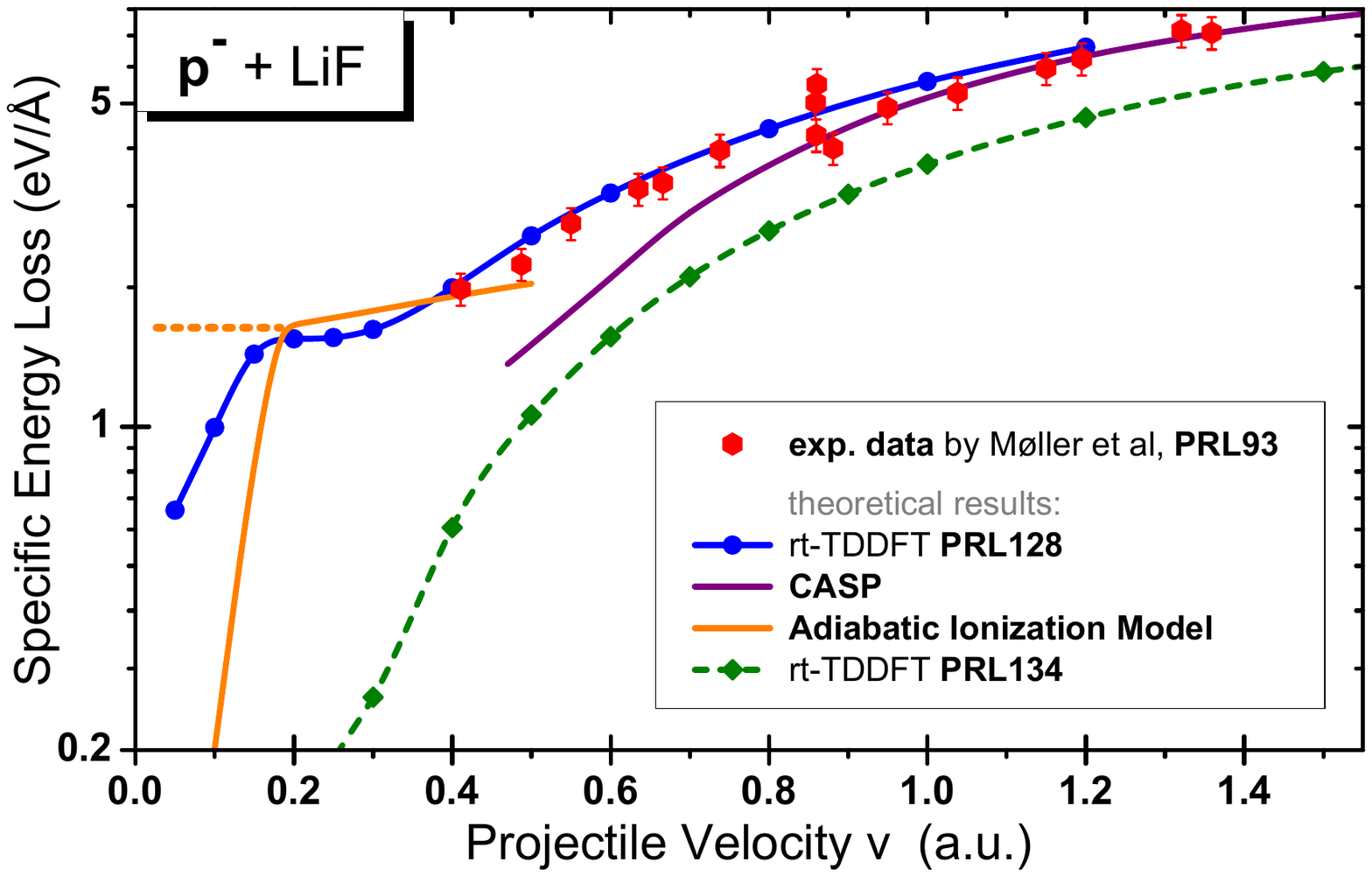}
    \caption{LiF stopping power vs. $\bar p$ velocity. 
Red symbols: Corrected CERN data~\cite{PRL93} (see text). 
Curves: CASP theory (purple)~\cite{CASP}, results of two rt-TDDFT codes (blue and green) and the AIM (orange) discussed above.}
    \label{fig:F3}
\end{figure}

Fig. \ref{fig:F3} presents specific energy losses on a logarithmic scale for $\bar p$ in LiF at low velocities, below the stopping-power maximum. 
We have scaled the displayed experimental data by Møller et al.\ \cite{PRL93} for $\bar p$ by a factor of 1.13 to align 
their proton data with the most precise recent proton measurements (for further details see \cite{PRLComment}). 
Such a possible systematic deviation was already noticed in their original publication and might be due to the 
observed grain structure of LiF or due to the half-covered target foil preparation 
(a thin carbon foil half with and half without evaporated LiF for simultaneous reference measurements).\cite{PRL93} 
Additionally, the graph includes four different theoretical results as introduced below:

\begin{itemize}
 \item CasP stopping-power results~\cite{CASP} for $\bar p$ are shown as a purple curve. These calculations have been performed for the LiF compound in the unitary-convolution approximation (UCA) mode, with shell correction and Barkas binary option. At high velocities down below the stopping-power maximum, these calculations should be highly reliable\cite{PRLComment} as they include some important additional terms beyond $1^{st}$ order perturbation theory, the Barkas term (proportional to the particle charge $q^3$) and a series of even powers of $q$, corresponding to the Bloch or strong-potential Born theory. At still lower velocities, however, perturbative expansions should consider higher odd and even powers of $q$ and this limits the range of validity and explains the deviation of CasP results from the experimental $\bar p$ data below about $v \approx 1 a.u.$.

 \item The solid blue curve shows results of a so-called real-time Time-Dependent Density-Functional Theory (rt-TDDFT).\cite{PRL128} 
These non-perturbative calculations use weighted impact-parameter averaging of antiproton trajectories along the $\langle 111\rangle$ direction. 
Comparison with the $\langle 001\rangle$ case as well as with different cluster sizes and impact parameter steps indicate numerical uncertainties around 5\%. 
To within this uncertainty there is perfect agreement with the experimental data and at high particle velocity also with the CasP results. 
This paper and the corresponding results are denoted PRL128 in the following.

 \item The dashed olive curve displays another rt-TDDFT solution, namely the PRL134 results\cite{PRL134}, the focus of our critique.\cite{PRLComment} 
These PRL134 calculations seem to employ a single trajectory along an incommensurable crystal direction. 
We agree that a huge set of non-channeling trajectories is somewhat closer to the experimental situation, 
but among other problems a single trajectory may lead up to 100\% error as is shown in the next main section of this paper.
The PRL134 authors claim that their results are superior to the PRL128 results (blue curve) which is obviously not consistent with Fig. \ref{fig:F3} 
and the deviation from the existing experimental data reaches even a factor of $3$ at low speeds. 

 \item The AIM energy-loss results depicted as a solid orange curve in Fig. \ref{fig:F3} follows directly 
from eq. \ref{eq:SeFormula} and the critical impact parameters are derived from curve crossings in Fig. \ref{fig:F2} 
between the excitation threshold as well as various collisional-broadening curves. 
Contrary to the CasP range of validity, the adiabatic ionization model should be reliable at low velocities, where non-adiabatic interaction terms are very small. 
Below a speed of 0.5 a.u., the AIM agrees reasonably well with the existing experimental data and with the PRL128 calculations.
\end{itemize}

As the contradictory statements in PRL134 are mainly concerned with the low-velocity behavior 
of $\bar p$ stopping powers in LiF, we will restrict the remaining discussion of Fig. \ref{fig:F3} on this part only.
The PRL128 results (blue curve) involve an interesting knee-like structure below a velocity of $0.3~a.u.$. 
A similar but somewhat sharper knee structure is also visible in the AIM results and this is easily understood. 
This AIM knee relates directly to the impact parameter 
dependence of the blue excitation-threshold curve in Fig. \ref{fig:F2}.
The steep part of the threshold curve above $b=2~a.u.$ leads to the nearly flat AIM results above $v=0.2~a.u.$
and the plateau in Fig. \ref{fig:F2} yields the steep AIM drop-off towards low speeds in Fig. \ref{fig:F3}.  

In a more advanced AIM solution outlined above, the step-function assumption 
for the AIM collisional broadening could be replaced by a smooth probability distribution.
In this case, the AIM results are expected to come closer to the knee structure 
of the blue PRL128 curve at low speeds. 
An even more striking result is the possible constant energy loss down to
the antiproton capture limit, shown by the dotted orange line.
This line is based on the assumption that the plateau in Fig. \ref{fig:F2} 
is actually consistent with a vanishing gap below a distance of $2~a.u.$, 
equivalent to a singly occupied F$^-$-2p orbital merging with the LiF conduction band.
Thus, this dotted orange line is fully consistent with predictions by Fermi and Teller~\cite{FermiTeller}
and with the vanishing gap as assumed by Solleder et al.~\cite{MCSCF}. 
All these results are inconsistent with results and statements in PRL134~\cite{PRL134} 
as is clearly indicated by the tremendous deviations of the green dashed curve 
from experimental data (red symbols), from AIM results (orange curves) and 
from the blue rt-TDDFT curve extracted from PRL128~\cite{PRL128}. 
Possible explanations for the disagreement of PRL134 from well-accepted literature and 
quantitative results outside their own group are discussed in the following section.

\section{Additional uncertainties in the  Phys.Rev.Lett. 134 paper \cite{PRL134}}
\label{sec:Errors}

In the following, we highlight specific statements in the PRL134 paper that appear to be incorrect or, at the very least, require further clarification.

\subsection{On capture and resonance effects for antiprotons in a LiF crystal}
\label{subsec:resonance}

\begin{itemize}
 \item The statements on the antiproton capture in the paper seem to make no sense. 
Orbiting of $\bar p$ around an F nucleus is of course inconsistent with the straight-line trajectories used in both TDDFT calculations~\cite{PRL128,PRL134}. 
Experimentally, it is also meaningless since captured antiprotons would most likely escape their detection. 
As discussed by Fermi and Teller~\cite{FermiTeller} for $\pi ^-$, 
capture is only likely when the total kinetic energy is comparable 
to the undistorted gap energy corresponding to $v_{capture} \approx 0.03~a.u.$ for $\bar p$.
Hence, if this capture statement is an argument for selecting only 
trajectories with large distances to $F^-$ ions, this would be wrong from all perspectives.

 \item The so-called Horsfield model 
(defined by eq. 1 and displayed as a fit curve in Fig.3 of PRL134) 
considers the interaction of a channeled particle with 
a de-localized state close to the center of the gap. 
No other publication known to us reports or assumes 
the existance of such a state in a pristine LiF crystal structure.
Valence-exciton energies, e.g., are far above the center of the gap and furthermore, 
the Frenkel-type exciton structure is pinned to the original $F^-$ ion.~\cite{Exciton}
Antiprotons, on the other hand, experience individual and localized energy losses, 
clearly separated from other interaction sites and events in space and time. 
Specifically at small impact parameters and low velocities, 
electron promotion of a singly occupied 2p orbital at an $F^-$ ion 
may either lead to conduction-band population (with a finite energy loss) 
or the orbital evolves back into an undisturbed 2p state at 
the host F nucleus during the departure of the antiproton.~\cite{MCSCF} 
The latter case results in zero energy loss, and because there is 
no memory of the previous interaction event, further couplings or even resonances are practically impossible.
The Horsfield model could thus only be applied to ultra-dense $\bar p$ pulse packets 
where the spatial structure of the $\bar p$ packet is coherent with the LiF lattice structure. 
This, however, is a very unrealistic scenario. 

A completely different argument against the application of the 
Horsfield model is its direct dependence on resonance conditions of 
a periodic atom structure in a crystal (see eq. 1 of PRL134).
In the Supplemental Online Material of the PRL128 paper\cite{PRL128}, one can find a comparison of 
$\bar p$ stopping powers for motion along the $\langle 111\rangle$ and $\langle 001\rangle$ directions. 
These results indicate that there is no strong dependence on the direction of motion and thus on the periodicity. 
Furthermore, our AIM results, do not depend on any direction at all 
but agree resonably well with the measured antiproton data at low speeds. 

 \item One main point of PRL134, is the criticism regarding TDDFT time integration 
along highly symmetric trajectories (parallel to crystal directions) as used in PRL128~\cite{PRL128}. 
Given a large density of de-localized in-gap states and 
very specific electron-transport properties, different types of dynamic resonance effects might exist in principle. 
As discussed above, however, such states do not play a role in pure LiF 
(with or without $\bar p$ or $p$). 
Antiprotons, e.g., can excite valence excitons at large impact parameters~\cite{MCSCF} 
(or depopulate them at small impact parameters), but these excitons are practically 
immobile~\cite{Exciton} and cannot contribute to resonance effects.
The resonant coherent excitation of ions under channeling conditions~\cite{RCE-Datz} is well known 
to excite and/or ionize bound projectile states, leading to significant changes in the energy loss spectrum~\cite{RCE-Moak}. 
But, the repulsive $\bar p - e^-$ interaction prohibits bound 
projectile-electron states at an antiproton.
One may also speculate that the long ranged Coulomb interaction or collective effects 
may couple neighboring ionic $F^-$ states, 
e.g., via wake-potential oscillations behind an antiproton. 
At projectile velocities below the materials Fermi velocity, however, 
wake-potential oscillations do not exist and 
the electronic interactions are known to be of short range. 
The latter was stated for $\bar p$ in LiF some time ago in a paper of one member of the PRL134 author group:
"The projectile energy loss mechanism is observed to be extremely local" ~\cite{PRL99}.
\end{itemize}

\subsection{On the importance and accuracy of incommensurate trajectories}
\label{subsec:trajectories}

One key difference between the rt-TDDFT calculations in PRL \cite{PRL134} and \cite{PRL128} lies in the method used to determine the stopping power for a truly random direction, thereby minimizing directional effects. As discussed above, however, directional effects should play no significant role for $\bar p$ at low speed.
After carefully reviewing the manuscript, one cannot be sure, but PRL134 \cite{PRL134} seems to employ just a single trajectory along an incommensurable crystal direction. In principle, such a procedure eliminates the need for sampling over impact parameters. However, it introduces a statistical uncertainty that is not easily quantified as it depends on the details of the impact-parameter dependent energy loss $\Delta E (b)$.  $\Delta E (b)$, on the other hand, is connected to the projectile velocity and correspondingly to the energy loss mechanisms.

\begin{itemize}
 \item There is no fluctuation of the data points around a hypothetical smooth curve in Fig 3 of PRL134. 
Thus, a fixed set of only one or a few trajectories might have been used. 
This was not stated explicitly in the paper, and it is misleading for any reader outside that research group.

 \item Here we present a quantitative estimate of the statistical uncertainty for the PRL134 evaluation of an incommensurate trajectory with a path length of about 48 \AA ,~as shown in their Fig. 2. Guided by the AIM, we assume a simplified low-energy promotion mechanism that leads to a fixed energy loss below a specific distance-of-closest-approach and otherwise zero energy loss (at larger impact parameters).
In Figs. \ref{fig:F4} and \ref{fig:F5}, we present two histograms for the number distribution of such 
small-impact parameter events defined by the critical impact parameter $b_c = 2 ~a.u.$ between straight-line trajectories and the F atoms in LiF along a fixed $\bar p$ velocity vector. 
Each trajectory corresponds to a different random start position. 
Using a finite trajectory length $L \approx 48$ \AA, the F atom density $\rho _F = 0.06088$  atoms/ \AA$^3$ and a reaction cross section of $\sigma_{react}= \pi b_c^2$, we estimate a mean number of $N_{react} = L \cdot \rho _F \cdot \sigma_{react} = 10.3$ collisions per random trajectory. Within about 4\%, this agrees with the mean collision number determined from the $10^5$
calculated trajectories in Fig. \ref{fig:F4} (10.9 collisions) and in Fig. \ref{fig:F5} (10.6 collisions). 
These Figs, however, show very different shapes, although the two histograms 
differ only by the two fixed 3-dimensional trajectory directions. 
These directions are [1 $\phi$ $\phi^2$] as suggested by A.A. Correa~\cite{Correa}
(cited and perhaps used in PRL134) as well as [1 $\pi/3$ $e/2$], 
another possible incommensurate direction that we introduce only for comparison purposes. 
A typical trajectory with a random start 
position in Fig. \ref{fig:F5} involves a mean error of 12\%, 
which is below the expected counting statistics for 10.6 events. 
An evaluation of Fig. \ref{fig:F4}, however, yields a mean error of 43\% 
and the asymmetric number distribution extends from 0 ionization events (error = -100\%) 
to 16 ionization events (corresponding to +47\% error). Hence, for the investigated test 
conditions (adjusted to PRL134) the trajectory direction in Fig. \ref{fig:F4} based on the golden ratio~\cite{Correa} 
involves a very strong sensitivity on the random start coordinates. 
For a single trajectory, this procedure may thus lead to an extremely large overall error.
\begin{figure}[ht]
    \centering
    \includegraphics[width=0.4\linewidth,angle=0]{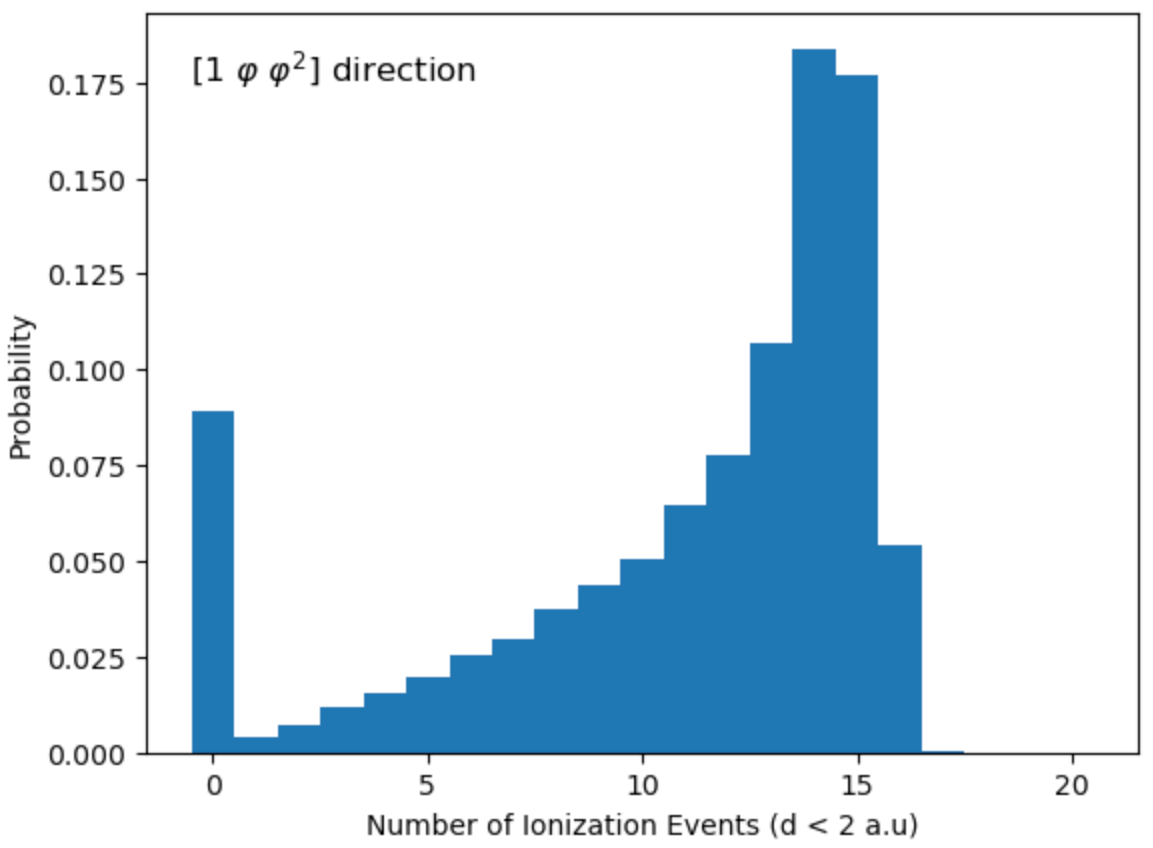}
    \caption{Histogram of the distribution of collision events with F atoms for a large set of straight-line trajectories along an incommensurate [1 $\phi$ $\phi^2$] direction using $\phi=(1+\sqrt 5)/2$ and random spatial offsets (see~\cite{Correa}).}
    \label{fig:F4}
\end{figure}
\begin{figure}[htb!]
    \centering
    \includegraphics[width=0.4\linewidth,angle=0]{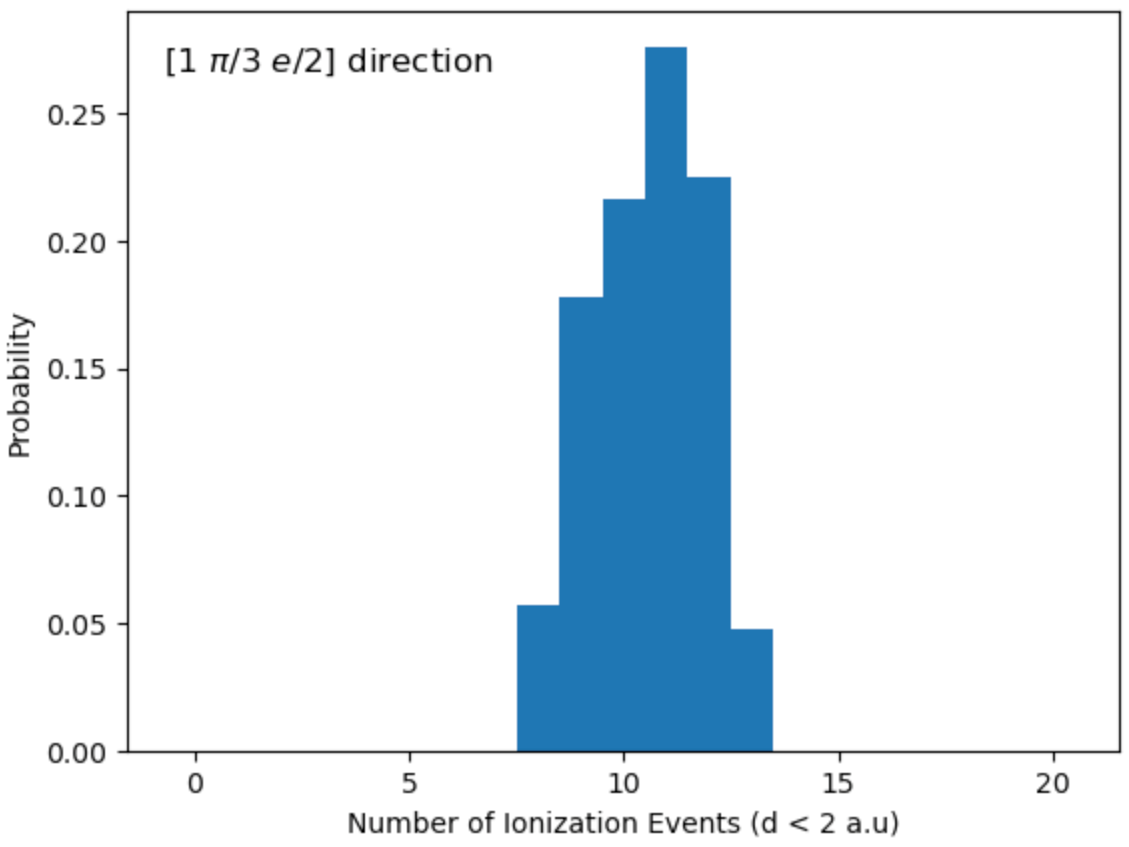}
    \caption{Histogram as in Fig. \ref{fig:F4}, but for the alternative incommensurate [1 $\pi/3$ $e/2$] direction.}
    \label{fig:F5}
\end{figure}

 \item In PRL134 it is stated that "the incommensurate-trajectory method has been shown to be the 
simplest and satisfactorily accurate", but the PRL134 authors cite only themselves and 
one external group that focuses on computational speed improvements via machine learning. 
Our above quantitative investigation of the reliabilty of the incommensurate-trajectory method 
for trajectories of finite length in a LiF lattice sheds new light on the accuracy question. 
For reaction cross sections determined by close collisions, about 10 or more 
of such trajectories are needed for obtaining results with a few-percent accuracy. 
The choice of a mathematically interesting incommensurate direction \cite{Correa}, 
however, can even worsen the accuracy problem. 
\end{itemize}

\subsection{On the methodology of the PRL134 quantum calculations}
\label{subsec:methodology}

\begin{itemize}
 \item Concerning the interaction potential, the PRL134 authors state 
"For the proton, a local Coulomb pseudopotential for hydrogen was used, 
with the equivalent repulsive pseudopotential for the antiproton." 
A pseudo potential is used to block specific bound states and/or to restrict some interaction terms. 
Exactly this might be responsible for the missing stopping-power contributions of 20\% at high velocities (if the singularity is removed) 
and partly for the 200\% at $v= 0.4~a.u.$ (if the long ranged dipole interaction is suppressed).

 \item So far we understand, do the PRL134 calculations 
follow the time dependence of $p$ and $\bar p$ induced excitations
along an incommensurate trajectory of about 48 \AA ~in length. Such a long trajectory improves 
the collision statistics and simplifies the linear regression for extracting stopping powers. 
However, this involves the disadvantage that the one 3x3x3 supercell is penetrated 
successively for a few times by the same particle (at different positions). 
Dependent on the properties of the supercell boundary  (such as simple periodicity or absorbing supercell surfaces), the supercell might not represent an undisturbed solid after a first passage. The valence-band population will change, and there might be an evolution of populated spurious conduction band-states that could build up chaotic continuum-wave structures or block further ionization events inside the supercell.  Both effects would degrade the accuracy of the final stopping-power results. 
\end{itemize}

\section{Conclusion}

In summary, our investigation has identified several noteworthy 
inconsistencies and misunderstandings in the PRL134 paper. 
In the light of these problems, we believe that PRL134 
should only be used as a reference with caution.

\bibliographystyle{ieeetr}

\end{document}